\documentclass[12pt]{article}
\usepackage{lineno}
\usepackage{graphicx}
\usepackage{amsmath}
\usepackage{hhline}
\usepackage{amssymb}
\usepackage{times}
\usepackage{cite}
\usepackage{array}
\usepackage{rotating}
\usepackage{multirow}
\usepackage{bigstrut}
\usepackage{color}
\definecolor{rltred}{rgb}{0.75,0,0}
\definecolor{rltgreen}{rgb}{0,0.5,0}
\definecolor{rltblue}{rgb}{0,0,0.75}

\newlength{\dinwidth}
\newlength{\dinmargin}
\begin{document}  

\newcommand{\GeV}{\rm G\eV}
\newcommand{\slowpi}{\pi_{\mathit{slow}}}
\newcommand {\gapprox}
   {\raisebox{-0.7ex}{$\stackrel {\textstyle>}{\sim}$}}
\newcommand {\lapprox}
   {\raisebox{-0.7ex}{$\stackrel {\textstyle<}{\sim}$}}
\def\gsim{\,\lower.25ex\hbox{$\scriptstyle\sim$}\kern-1.30ex%
\raise 0.55ex\hbox{$\scriptstyle >$}\,}
\def\lsim{\,\lower.25ex\hbox{$\scriptstyle\sim$}\kern-1.30ex%
\raise 0.55ex\hbox{$\scriptstyle <$}\,}
\newcommand{\trp}{p_{\perp}}
\newcommand{\trmm}{m_{\perp}^2}
\newcommand{\trpp}{p_{\perp}^2}
\newcommand{\alps}{\alpha_s}
\newcommand{\sqrts}{$\sqrt{s}$}
\newcommand{\LO}{$O(\alpha_s^0)$}
\newcommand{\Oa}{$O(\alpha_s)$}
\newcommand{\Oaa}{$O(\alpha_s^2)$}
\newcommand{\PT}{p_{\perp}}
\newcommand{\JPSI}{J/\psi}
\newcommand{\sh}{\hat{s}}
\newcommand{\uh}{\hat{u}}
\newcommand{\PO}{I\!\!P}
\newcommand{\dgr}{^\circ}
\newcommand{\pbarnt}{\,\mbox{{\rm pb$^{-1}$}}}
\newcommand{\WBoson}{\mbox{$W$}}
\newcommand{\fbarn}{\,\mbox{{\rm fb}}}
\newcommand{\fbarnt}{\,\mbox{{\rm fb$^{-1}$}}}
%
%
\newcommand{\et}{\ensuremath{E_t^*} }
\newcommand{\rap}{\ensuremath{\eta^*} }
\newcommand{\gp}{\ensuremath{\gamma^*}p }
\newcommand{\dsiget}{\ensuremath{{\rm d}\sigma_{ep}/{\rm d}E_t^*} }
\newcommand{\dsigrap}{\ensuremath{{\rm d}\sigma_{ep}/{\rm d}\eta^*} }
\newcommand{\ftwocc}{$F_2^{c\bar{c}}$}
\newcommand{\ftwobb}{$F_2^{b\bar{b}}$}
\newcommand{\msbar}{\mbox{\small {$\overline {MS}$}}}
\newcommand{\mbar}{\mbox{\small {$\overline {m}$}}}
\def\Journal#1#2#3#4{{#1} {\bf #2} (#3) #4}
\def\NCA{\em Nuovo Cimento}
\def\NIM{\em Nucl. Instrum. Methods}
\def\NIMA{{\em Nucl. Instrum. Methods} {\bf A}}
\def\NPB{{\em Nucl. Phys.}   {\bf B}}
\def\PLB{{\em Phys. Lett.}   {\bf B}}
\def\PRL{\em Phys. Rev. Lett.}
\def\PRD{{\em Phys. Rev.}    {\bf D}}
\def\ZPC{{\em Z. Phys.}      {\bf C}}
\def\EJC{{\em Eur. Phys. J.} {\bf C}}
\def\CPC{\em Comp. Phys. Commun.}

\begin{titlepage}

\begin{flushleft}
\end{flushleft}

\vspace{2cm}

\begin{center}
\begin{Large}
  
{\bf Comparison of Inclusive Charm and \\
 Beauty Cross Sections in \\
 Deep-inelastic Scattering at HERA \\
      with Theoretical Predictions}

\vspace{2cm}

\end{Large}
\end{center}

\begin{center}
P.~D.~Thompson $^{a}$ \\

\vspace{1cm}

$^a$ E-mail: pdt@hep.ph.bham.ac.uk, School of Physics, University of Birmingham, Birmingham B15 2TT, UK \\
\end{center}

\vspace{2cm}

\begin{abstract}
  \noindent The  measurements of inclusive charm and beauty
  cross sections in deep-inelastic scattering  
 $ep$ collisions at HERA are 
  compared with the predictions of 
  perturbative quantum chromodynamics from the CTEQ and MRST
fitting groups, employing a range of theoretical schemes. 
   The differences in the theoretical predictions 
are discussed and the theoretical uncertainties investigated.

\end{abstract}

\begin{center}
\end{center}

\end{titlepage}

\newpage

\section{Introduction}
\label{introduction}

The study of heavy flavour production in deep-inelastic
scattering (DIS) $ep$ collisions at HERA
provides a test of perturbative
quantum chromodynamics (pQCD).  
In particular, the presence of the heavy quark mass $M$ provides
an additional `hard' scale to the momentum
transfer of the exchanged boson $Q$.
The perturbative series has to be treated in different ways 
depending on the relative magnitude of $M$ and $Q$ and is, therefore,
a stringent test of the QCD factorization theorem.
The description of heavy flavour processes is particularly important
in precise global QCD analyses of DIS structure functions as heavy flavours
form an increasingly significant contribution to the total
cross section with increasing energy.  
For example, at values of $Q^2 = 650~{\rm GeV^2}$ charm (beauty) production
has been measured~\cite{H1f2cf2bhighq2} to be $\sim25\%$ ($\sim2.5\%$) 
of the total DIS cross section. 
The measurement of the charm and beauty structure 
functions, \ftwocc\ and \ftwobb\@, provides a more direct sensitivity 
to the gluon distribution of the proton than the inclusive structure 
function $F_2$.  The understanding of the gluon and quark distributions
in the region of low $x$ has important implications for the measurement
of standard model and new physics processes at hadron colliders such as the 
Tevatron and LHC.
Recent measurements of \ftwocc\ and \ftwobb\ have been made at 
HERA~\cite{H1f2cf2bhighq2,H1f2cf2blowq2} 
based on a technique using the displacement of tracks from the primary
interaction vertex, which arise from long lived charm and beauty hadrons,
and have small extrapolations to the inclusive phase space.
The measurements of \ftwocc\ using this technique are found to be compatible 
with those obtained using a largely independent method based on
the reconstruction of exclusive charm 
mesons~\cite{H1Dstar94,ZEUSDstar97,H1Dstar,ZEUSDstar},
giving increasing confidence that the extrapolations from the 
exclusive to the inclusive phase space, for the $Q^2$ range
of the displaced track measurements, are well controlled.

In this paper the status of the theoretical description of the 
latest published data and the theoretical uncertainties are evaluated.  
Firstly, 
the various theoretical treatments available 
for the description of heavy flavour production 
within pQCD are briefly introduced.  
Secondly, a comparison of the different  theoretical
schemes with the experimental data is made for \ftwocc\ and \ftwobb\ .
Thirdly, the uncertainties of the 
theoretical predictions for charm and beauty are compared.
In the last section the influence of the
gluon distribution on the cross sections is investigated in more detail.
These investigations aim to provide insight into the areas of 
phase space where future
measurements of heavy flavour data can contribute to best
improve the theoretical understanding.

\section{Theoretical Treatments of Heavy Flavour Production}
\label{heavyflavourtreatments}

In this section the three sets of the most common theoretical treatments 
of heavy quark production in DIS are briefly introduced. 
The first two methods attempt to reduce the dual scale problem
involving $Q$ and $M$ 
to an effective, and hence approximate, single scale problem.
The third scheme tries to unify the two different approaches,
which are valid at the extremes of phase space, to provide the most
accurate description over the whole kinematic range.
The sets of parton distribution functions (PDFs)
considered in this paper, which employ all three heavy flavour schemes,
are restricted to those from the two QCD fitting groups 
of CTEQ~\cite{Cteq5,Cteq65} and 
MRST~\cite{Martin:2004dh,ThorneNNLO,mrstff}.

\subsection{Fixed Flavour Number Scheme}
\label{ffns}
In order to correctly describe the production of a pair
of heavy flavour quarks around the threshold of $W = 2M$, where
$W$ is the photon-proton centre of mass energy, the mass of the heavy quark 
should be explicitely taken into 
account in the calculations.  
This is done in the fixed flavour number scheme (FFNS) where
the heavy flavour quarks are treated as `heavy' particles,
never as massless partons.  The leading order production process for heavy
flavour in this scheme is based on the `massive' boson-gluon fusion (BGF) 
matrix element to order $\alpha_s$, generated mainly from the
gluon distribution of the proton. The BGF process is one order higher in
$\alpha_s$ than the lowest order quark parton model (QPM) contribution for 
the production of light quarks. 
The maximum number of flavours $N_f$ contributing to the strong
coupling evolution is limited to the maximum number of `light' or 
massless flavours.  In the case where both charm and beauty 
are treated as massive $N_f=3$.  In the case where charm is 
treated as massless and beauty as massive then $N_f=4$.
In this paper only those FFNS PDFs for which $N_f=3$ are considered.
The heavy flavour FFNS BGF-like coefficients have been calculated to
next-to-leading order (NLO) i.e. to order $\alpha_s^2$~\cite{ffnsnlo}.

The FFNS provides
the most reliable description of heavy flavour production
around the kinematic 
threshold~\footnote{The threshold of  $W = 2M$ corresponds to 
$Q^2 = 4M^2 x/ (1-x)$.}
(in the absence of a non-perturbative contribution).
However, the presence of mass dependent terms of the form $\ln Q^2/M^2$, 
which are kept in the expansion for all values of $Q^2$, spoils the
convergence of the series as $Q^2 \rightarrow \infty$.
QCD does not predict the scale at which the  terms 
become infra-red unsafe and, as will be seen, the FFNS provides
an adequate description of the existing experimental HERA data
over the entire kinematical range.
However, theoretically it remains prudent to take
these divergent terms into account. This is dealt with by the variable
flavour number schemes described in the next sections.
Another disadvantage of the FFNS is that the hard-scattering
matrix elements are generally an order of $\alpha_s$ higher
than the `massless' calculations and have not been calculated
for many processes, for example, charged current scattering.
This is particularly problematic for global QCD analyses
where many of the standard data sets which are used in the
fits have to be omitted in a FFNS fit, resulting in less well
constrained partons.

The FFNS is implemented by the CTEQ and MRST groups for either
three (CTEQ5F3~\cite{Cteq5}, MRST2004FF3~\cite{mrstff}) 
or four (CTEQ5F4~\cite{Cteq5}, MRST2004FF4~\cite{mrstff}) 
flavours of light quarks.  
The limited applicability of the FFNS is the reason for the
absence of FFNS PDFs in the most recent
set of CTEQ parton densities (CTEQ6.5~\cite{Cteq65}) i.e. there are no
CTEQ6F3 or CTEQ6F4 PDFs.
However, parton sets compatible with the FFNS are still essential
for making predictions for the hadronic final state in $ep$ collisions
since the main NLO program HVQDIS~\cite{HVQDIS}, which implements the
massive BGF matrix elements to NLO and provides differential 
cross sections for the heavy quarks, is calculated in this scheme. 
The MRST2004FF partons have been produced specially for this purpose 
from another set of PDFs which used a more general scheme for heavy
flavour production (see section~\ref{vfns}).  It should
be noted that because the MRST2004FF 
PDFs are not extracted from a FFNS fit themselves, they generally
provide a poor description of the inclusive proton structure function 
$F_2$.

As was highlighted in~\cite{mrstff} the original calculations
of the massive NLO coefficient functions
were evaluated for the fixed flavour definition of the strong
coupling $\alpha_s^{N_f=3,4}$.  Therefore,
when fitting at NLO in the FFNS a compatible fixed flavour
definition of $\alpha_s$ should be used.  For the H1~\cite{h1lowq2} 
and ZEUS~\cite{zeusqcd} FFNS fits the more commonly used variable 
five flavour definition of $\alpha_s^{N_f=5}$ was used.
The incompatibility of coefficients and schemes leads to too
much evolution at low $Q^2$ and too little evolution at higher
$Q^2$, causing inaccuracies for the predictions of
heavy flavour cross sections. 
In order to use a value of $\alpha_s^{N_f=3,4}$ compatible with the
world average value of $\alpha_s^{N_f=5}(M_Z) \sim 0.118 $ in a FFNS fit,
the couplings (via the value of $\Lambda^{QCD}$) should be matched 
at some low scale (typically $m_c$), so that it is possible to describe the 
bulk of the experimental data that are in the low $Q^2$ region.
Note that when the fixed three-flavour definition of $\alpha_s^{N_f=3,4}$
is evolved to $M_Z$ it will have a
low value $\alpha_s(M_Z) \sim 0.1$ when compared with the world average.
The correct FFNS implementation was used by H1 in a recent fit to
inclusive diffractive data~\cite{diffcharm}. 
The ZEUS collaboration use the
Thorne-Roberts scheme~\cite{VFNS2}(see section~\ref{vfns}) 
to account for heavy flavour mass effects in their standard 
PDFs~\cite{zeusqcd,zeusjets}.

\subsection{Zero Mass Variable Flavour Number Scheme}
\label{zmfns}
The simplest way to resum the divergent
$\ln Q^2/M^2$ terms in the perturbative expansion 
is to absorb them into the
parton distribution functions by treating them as
massless partons in the standard way.  
In the zero mass-variable flavour number scheme (ZM-VFNS) 
the heavy flavour particles are treated 
as massless partons.  
The heavy flavour partons contribute to the cross section
and evolution of $\alpha_s$ only when the factorization 
scale $\mu_f$ is larger than
some threshold value, usually chosen as $M$.
Since in the ZM-VFNS $Q$ is often used as the 
factorization scale, the heavy flavour parton
distributions are zero below $Q^2 < M^2$ and all other 
parton distributions are continuous across the
threshold $Q^2 = M^2$~\footnote{This holds to NLO but at NNLO
the heavy flavour parton distributions may be none zero below
$Q^2 = M^2$ and the other parton distributions may be
discontinuous across the mass thresholds.}.
In the ZM-VFNS
the heavy flavour parton densities provide the correct
theoretical behaviour when $Q^2$ is large ($Q^2 \gg M^2$) 
but the approximation $M=0$ deteriorates as $Q^2$ 
becomes the same order of magnitude as $M^2$.   
To improve the applicability of the ZM-VFNS at low $Q^2$
more general schemes have been developed and are introduced 
below.

\subsection{General Mass Variable Flavour Number Schemes}
\label{vfns}
Whilst neither the FFNS nor the
ZM-VFNS can provide individually a satisfactory theoretical
description of heavy flavour production over the
whole kinematic range, the most reliable predictions 
can be obtained by combining the two and utilising
the most appropriate scheme at a particular $Q^2$.
The first scheme to attempt to unify the perturbative 
reliability of the ZM-VFNS at
large $Q^2$, whilst introducing a finite quark mass 
around threshold to correctly describe the production of heavy
flavours at low $Q^2$, was developed
by the ACOT~\cite{ACOT} group of collaborators.  
In simple terminology,
the DIS cross section in the VFNS\footnote{The formalism is often
also referred to as the general mass (GM) VFNS to help
distinguish it from the ZM-VFNS.} may be considered
as the sum of the fixed flavour and massless terms 
and a term which represents the overlap between the two, which
needs to be subtracted in order to avoid double counting of the 
cross section.  
Although all theoretical
groups adopt many of the underlying ideas of the 
ACOT scheme there remains some freedom in the
prescription for treating the scattering terms in different orders 
of $\alpha_s$ across the heavy flavour mass thresholds.   

The most recent set of standard PDFs from CTEQ (CTEQ6.5) is based on an
implementation of the general mass ACOT scheme that incorporates the
ACOT($\chi$)~\cite{ACOTchi} rescaling variable for a more accurate treatment 
of the kinematics, the S-ACOT~\cite{SACOT} prescription for 
a simplified treatment of 
the Wilson coefficients and other features, as described in~\cite{Cteq65}.

The MRST group adopt the Thorne-Roberts (TR)~\cite{VFNS2}
prescription for the VFNS.  
The TR VFNS is based on the ACOT scheme and implements the same
heavy quark coefficient function choice of the ACOT($\chi$) prescription 
for \ftwocc\ and \ftwobb\ but differs in other
choices. This scheme is used in the
standard MRST2004 set of parton distributions.  The TR VFNS 
was recently extended to NNLO~\cite{ThorneNNLO}.  As the order $\alpha_s^3$
scattering coefficients for massive heavy flavour production
have yet to be calculated, approximations,
that were found to be successful in estimating NNLO effects in the
inclusive cross section, are made. For this paper an update of the
NNLO partons compared to those originally appearing in~\cite{ThorneNNLO}
are used.  

As discussed in section~\ref{ffns} the most recent heavy flavour 
$ep$ final state calculation program available HVQDIS still uses 
the FFNS and, therefore, comparison with the 
results from the latest VFNS PDFs from CTEQ and MRST 
(with the exception of the specially generated MRST2004FF sets)
is difficult.

\subsection{Summary of PDFs and Parameters}
\label{theorysummary}
In this section the main QCD schemes for heavy flavour
production have been introduced. The PDFs from CTEQ and MRST which
are used in this paper are summarised in table~\ref{table:params}.   
As with all QCD calculations there are a number of choices 
in the parameters and scales used.  The table shows the values used
in the original extraction of the PDFs by the fitting groups. 
The effect of the variations of these parameters is investigated in 
section~\ref{theoryuncert}.

The masses of the heavy
quarks are important in the predictions 
and are shown in table~\ref{table:params}.
The values for CTEQ and MRST are $5 - 10\%$ different. 
The quark masses in the table are the 
`on-shell' mass or pole mass~\cite{PDG2006} and
are, therefore, slightly larger than the latest values in the
$\msbar$ scheme from the PDG~\cite{PDG2006} of 
$\mbar_c = 1.25 \pm 0.09 ~{\rm GeV}$ and 
$\mbar_b = 4.20 \pm 0.07~{\rm GeV}$. 
In comparisons of theoretical predictions
with final state measurements the `on-shell' mass values and
variations typically used are  $m_c=1.5\pm 0.2 ~{\rm GeV}$ and
$m_b = 4.75 \pm 0.25 ~{\rm GeV}$~\cite{hera-lhc}.

Other important parameters for heavy flavour
production are the factorization $\mu_f$ and renormalization $\mu_r$
scales. The former sets the scale at which the 
parton distribution is sampled and the latter the
scale at which $\alpha_s$ is evaluated.
In the case of light quark production $\mu_f$ and $\mu_r$ are usually set
to $Q$.
The presence of a `heavy' quark creates an additional 
uncertainty in the choice of the perturbative scale.  The perturbative scale 
in the expansion of the heavy flavour coefficients $\mu_M$ may be different 
to that chosen for the light quarks.
For the MRST PDFs $\mu_M$ is chosen to be the same as for the 
light quarks $\mu_M = Q$.  For the most recent PDF from CTEQ (CTEQ6.5) the
scale for heavy flavour is chosen as $\mu_M = \sqrt{Q^2+M^2}$.
As discussed in~\cite{Cteq65} the heavy flavour structure
functions \ftwocc\ and \ftwobb\ are theoretically 
not infra-red safe beyond NLO. However, the effects are expected to be small
and comparison with experimental results is still possible, although the 
NLO predictions may be sensitive to the choice of parameters such 
as $\mu_M$ (see section~\ref{theoryuncert}).

\begin{table}[h]
\begin{center}
\begin{tabular}{|c|c|c|c|c|c|}
\hline
\multicolumn{6}{|c|}{Summary of PDF Schemes and Default Parameters}\\
\hline
PDF & Order & Scheme & $\mu_M$ & $m_c~({\rm GeV})$ & $m_b~({\rm GeV})$ \\ \hline
CTEQ5F3 & $\alpha_s^2$ & FFNS & $Q$ & 1.3 & 4.5 \\
MRST2004FF3 & $\alpha_s^2$ & FFNS & $Q$ & 1.43 & 4.3 \\  
CTEQ6.5 & $\alpha_s^2$ & VFNS  & $\sqrt{Q^2+M^2}$  & 1.3 & 4.5 \\
MRST2004 & $\alpha_s^2$ & VFNS  & $Q$ & 1.43 & 4.3 \\ 
MRST2004 NNLO & $\alpha_s^3$ & VFNS & $Q$ & 1.43 & 4.3 \\
\hline
\end{tabular}
\end{center}
\caption{Summary of the theoretical schemes and default parameters 
used for comparison with the data in this paper.}
\label{table:params}
\end{table}

A further consideration in the comparison of the heavy flavour PDFs
with the HERA data is the data sets that were used as input 
in the analysis to extract the PDFs. The CTEQ and MRST PDFs are based 
mainly on the measurement
of the inclusive neutral and charged current cross sections at HERA,
inclusive fixed target data
and additional constraints on the high $x$ gluon are obtained from jet 
measurements at the Tevatron.  
The CTEQ6.5 PDF also includes the 
HERA heavy flavour data 
sets~\cite{ZEUSDstar97,H1Dstar,ZEUSDstar,H1f2cf2bhighq2,H1f2cf2blowq2}, and
the MRST2004 PDFs include HERA \ftwocc\ data~\cite{H1Dstar94,ZEUSDstar97}.

\section{Comparison with Experimental Data}
In this section the experimental data
on inclusive charm and beauty production are compared with the PDFs 
from MRST and CTEQ introduced in section~\ref{heavyflavourtreatments}.  
A brief summary of the experimental
methods employed to extract the data is also given.
The comparison aims to highlight those areas where
the theoretical predictions significantly differ.

\subsection{Inclusive Charm Production}

\subsubsection{Experimental Measurements}

Measurements of the inclusive charm structure function
$F_2^{c\bar{c}}$ at HERA were first performed using the reconstructed
decay products of $D$ 
mesons~\cite{H1Dstar94,ZEUSDstar97,H1Dstar,ZEUSDstar}.  The measured visible 
cross sections of the $D$ meson are then extrapolated to the full phase 
space using theoretical models.  At the lowest values of $Q^2$ 
the extrapolations can be
large ($\sim 3-5$~\cite{ZEUSDstar}) since the hadronic final state has
a low probability to be in the restricted angular and
transverse momentum ranges in which the measurements are made.
In particular, experimental cuts on the transverse momentum of the $D$ 
meson are required to be high in order to suppress combinatorial background.
The size of the extrapolation and, hence, the uncertainty on the extrapolation
reduces with increasing $Q^2$ as the hadronic final state recoils with a
larger transverse momentum. At large $Q^2$ ($Q^2=500~{\rm GeV^2}$) the 
extrapolation factor is
much smaller in the range $\sim 1.6-2.4$~\cite{ZEUSDstar}.  
The theoretical uncertainty
on the extrapolation was investigated in \cite{H1Dstar} by using 
an alternative model based on CCFM evolution~\cite{ccfm2}.  The
differences were found to be as large as $20\%$ at low $Q^2$ and $x$, 
reducing to around $5\%$ at higher $Q^2$.
In this paper the $D$ meson data shown from H1 and ZEUS are taken from the 
most recent publications~\cite{ZEUSDstar, H1Dstar} based on the 
reconstruction of $D^*$ mesons.
The experimental values of the charm structure function were calculated
using the FFNS and the program HVQDIS to evaluate the extrapolations
to the full phase space.  
In this procedure the predictions of HVQDIS for the differential 
$D^*$ meson cross sections are obtained by applying hadronisation
corrections, estimated using Monte Carlo simulations and parameterisations
of the $D^*$ fragmentation function~\cite{peterson}, 
to the parton level predictions.

Recently measurements~\cite{H1f2cf2bhighq2,H1f2cf2blowq2} 
of the charm cross section
have been made based on a technique which utilises the 
long lifetime of heavy flavour hadrons by measuring the displacement 
of tracks from the primary vertex. This allows access to 
lower momentum and a wider angular range than the reconstruction of
exclusive charmed mesons.  Provided the measurement is made in a
range of $Q^2$ where the hadronic final state receives a large
transverse momentum the technique can be used to measure the
inclusive charm and beauty cross sections.
The results from the displaced track
and $D$ meson analyses are 
found to be very compatible~\cite{H1f2cf2bhighq2,H1f2cf2blowq2}. 
It should be noted that although the extrapolation to the full phase 
space is reduced using the displaced track technique the combined 
statistical and systematic errors of the two methods,
which arise from different sources, are of similar size. 

\subsubsection{Comparison of Experimental Charm Data with PDFs}

\begin{figure}[htb]
 \begin{center}
\includegraphics[width=1.00\textwidth]{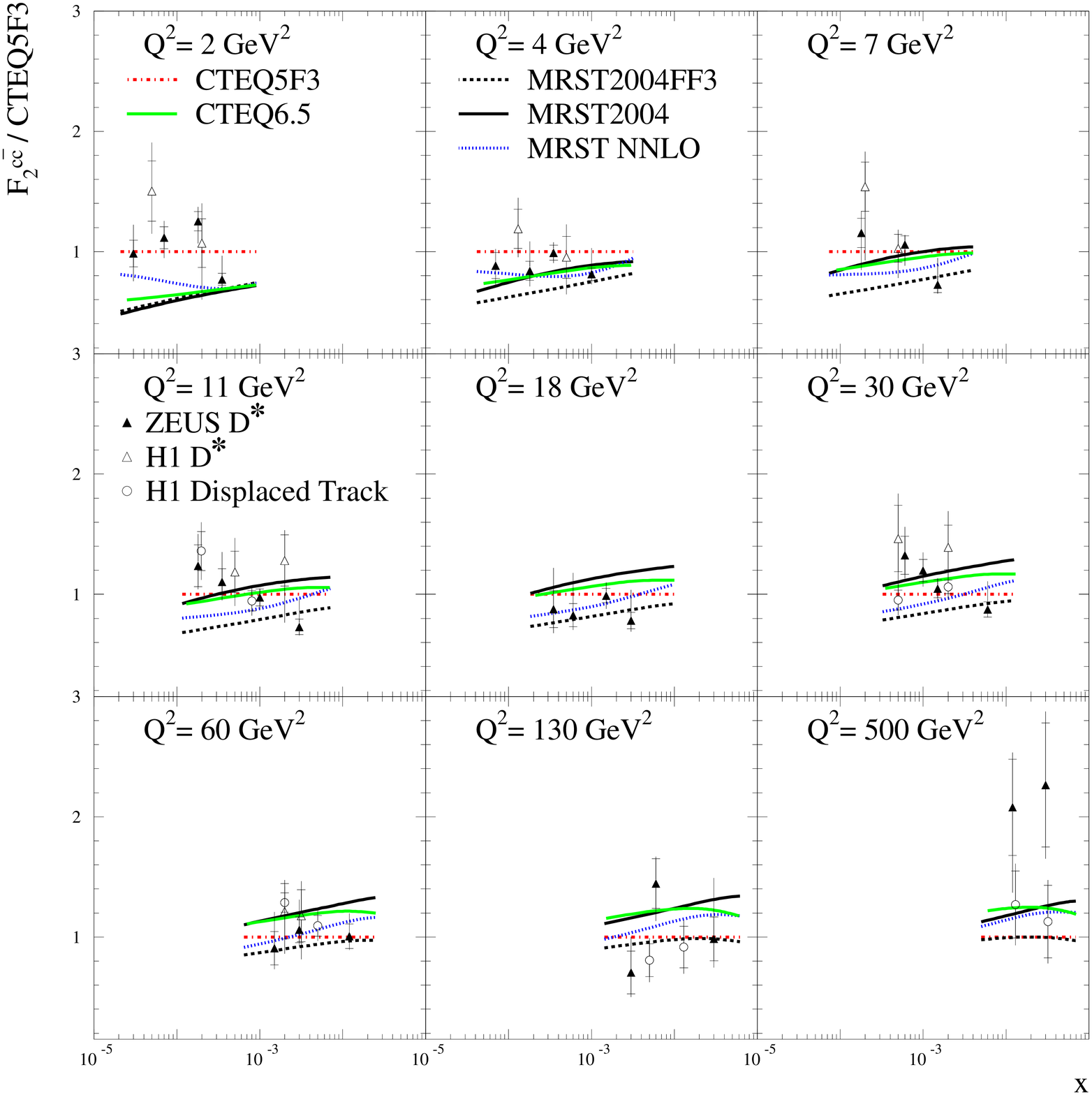}
   \caption{The ratio of the charm structure function \ftwocc\ 
to the prediction of CTEQ5F3 shown 
as a function of $x$ for different values of $Q^2$.
The data obtained using reconstructed $D^*$ mesons and the 
displacement of tracks from the primary vertex are shown
as points. The inner error bar corresponds to the statistical
uncertainty and the outer error bar to the statistical and systematic
errors added in quadrature.  The data are compared with
the theoretical predictions of perturbative QCD using different prescriptions
for the heavy flavour treatment from the MRST and CTEQ fitting 
groups.}  
\label{fig:f2c}
\end{center}
\end{figure}

Figure~\ref{fig:f2c} shows the ratio of the charm structure function \ftwocc\
to the FFNS prediction of CTEQ5F3 
as a function of $x$ for different values of $Q^2$ using the ZEUS $D^*$ binning
scheme.  
The data from the $D^*$ meson and displaced track methods are shown
as the points with errors.   
The H1 data points are interpolated to the
nearest bin centre in $Q^2$ using a parameterisation of  
the cross section from a H1 FFNS NLO   
QCD fit as used in~\cite{H1f2cf2blowq2}. 
The \ftwocc\ measurements contain a small QCD correction from the effect of 
the longitudinal cross section. 
To avoid this correction in future the reduced cross section should be measured
for heavy flavour production as was done in~\cite{H1f2cf2blowq2} 
and as has been the standard for the measurement
of the inclusive DIS cross section for a longer time. 

The data are compared with predictions using implementations of
the NLO FFNS (CTEQ5F3, MRST2004FF3), the NLO VFNS (CTEQ6.5, MRST2004) and a
NNLO VFNS (MRST2004 NNLO).  Each of the three groups of 
predictions are discussed below.

The CTEQ FFNS PDF (CTEQ5F3) provides a 
good description of the data throughout the $Q^2$ range, in particular at 
low $Q^2$, supporting the hypothesis that the FFNS provides the correct
description of heavy flavour production in this region. 
The MRST FFNS PDF (MRST2004FF3) is $50-30\%$ lower than the CTEQ FFNS PDF
at the lowest $Q^2$ value.  This is a consequence of 
the matching of the FFNS partons to those of the VFNS at 
$Q^2 = m_c^2$ and the MRST gluon distribution being smaller than CTEQ
in the region of small $x$ and $Q^2$. 
The two FFNS PDFs show increasingly similar predictions 
with increasing $Q^2$ and are almost identical at $Q^2 = 500 \ {\rm GeV^2}$.

The two NLO VFNS PDFs (CTEQ6.5, MRST2004) show similar
predictions of \ftwocc\ throughout the kinematic range of the figure.
They both exhibit a shallower $x$ dependence 
and a different $Q^2$ evolution than the CTEQ FFNS PDF
and are  around $40\%$  lower than the CTEQ FFNS
prediction at the lowest values of $Q^2$ and around $20\%$ higher at
the highest $Q^2$ values. 
In the low $Q^2$ region it may be expected that a
VFNS PDF tends towards the FFNS result. This does not 
appear to be the case for the CTEQ PDFs which 
were extracted using different heavy quark factorization
scales (see table~\ref{table:params}) and 
different DIS data sets, as discussed in~\cite{Cteq65}.  
As stated above the 
MRST FFNS PDF is matched to the VFNS NLO PDF at low values of $Q^2$.

The matching of the MRST and FFNS partons 
at low $Q^2$ allows to investigate
the difference in the evolution of the FFNS and VFNS partons with $Q^2$.   
The different evolution leads to a divergence, with the VFNS partons evolving
more quickly than the FFNS partons, with increasing values 
of $Q^2$ with a convergence again at the highest $Q^2$. As discussed
in~\cite{mrstff} this is attributed to missing terms in the FFNS evolution.
A similar behaviour was observed in an earlier comparison of FFNS and VFNS 
predictions for charm production~\cite{Chuvakin:2000zj}.

The NNLO VFNS PDF (MRST2004 NNLO) improves considerably 
the description of the charm data at the lowest $Q^2$ values 
when compared with the MRST NLO VFNS PDF.  
For the rest of the phase space the NNLO VFNS prediction
mostly lies between those of the MRST NLO VFNS and FFNS, only 
rising to the VFNS prediction at $Q^2 = 500 \ {\rm GeV^2}$.  

For the low $Q^2$ and low $x$ charm data, the good description provided
by the CTEQ NLO FF PDF compared with the need for higher orders
to improve the description by MRST indicates that this region is 
sensitive to the differences in the gluon distribution, heavy flavour scheme 
and the order of $\alpha_s$ used in the calculations.
For example, the ZEUS-JETS NLO PDF~\cite{zeusjets}, which implements the 
Thorne-Roberts VFNS for the treatment of heavy flavours, 
gives predictions for charm (not shown) which lie between the MRST2004 and 
CTEQ5F3 PDFs for the three lowest $Q^2$ bins.  The difference with respect to 
MRST2004 arises solely from the ZEUS-JETS gluon distribution being larger than 
MRST2004 in this region, whereas the difference to CTEQ5F3 is due to 
differences in both the scheme and gluon.

\begin{figure}[htb]
 \begin{center}
\includegraphics[width=1.00\textwidth]{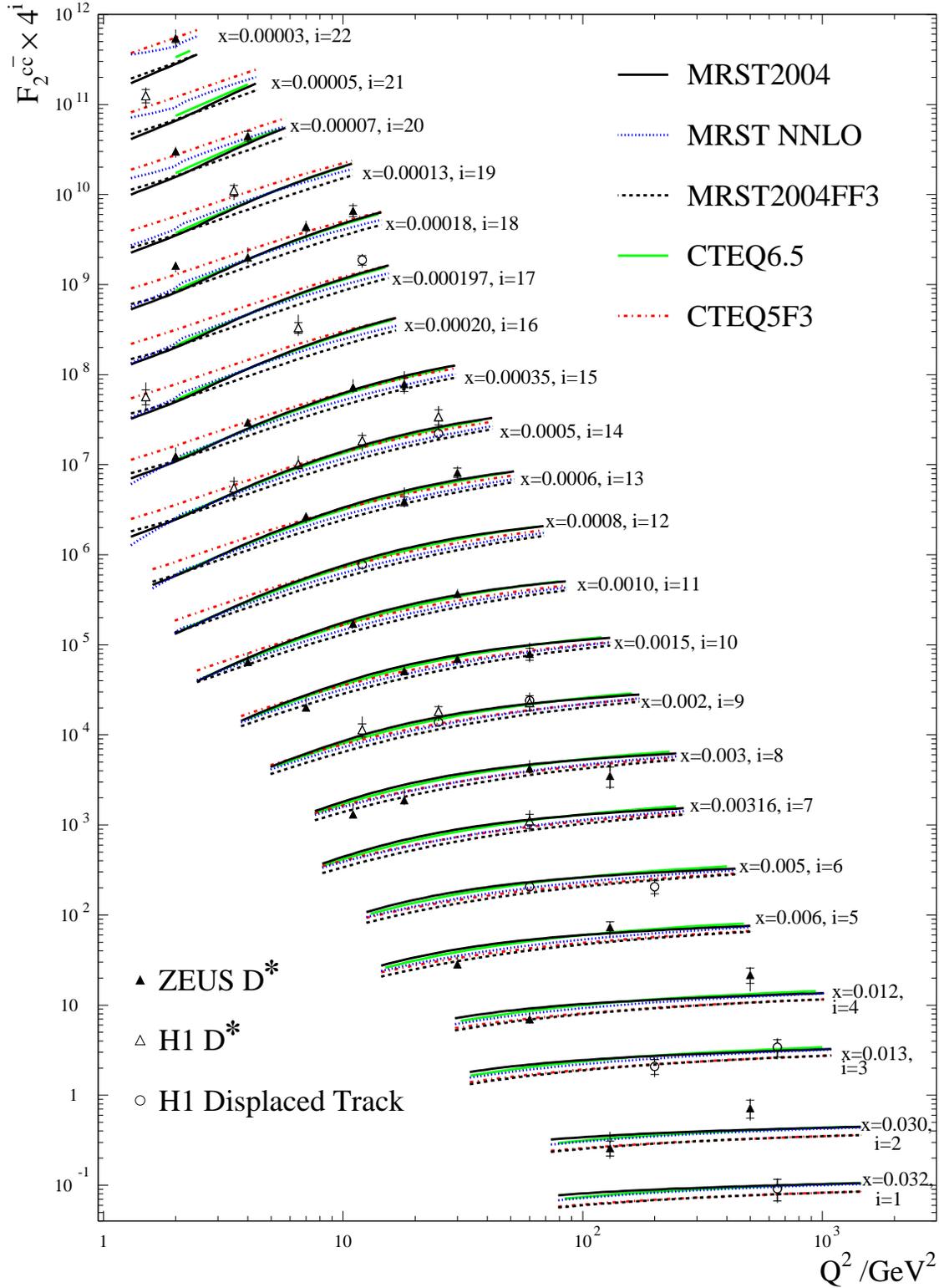}
   \caption{The charm structure function \ftwocc\
as a function of $Q^2$ for different values of $x$.
The data obtained using reconstructed $D^*$ mesons and the 
displacement of tracks from the primary vertex are shown
as points. 
The data are compared with
the theoretical predictions of perturbative QCD using different prescriptions
for the heavy flavour treatment from the MRST and CTEQ fitting 
groups.}  
\label{fig:scalingc}
\end{center}
\end{figure}

The data and theoretical predictions are compared again in 
figure~\ref{fig:scalingc} 
this time as a function of $Q^2$ for different values of $x$.
The charm data show positive scaling violations for the majority of the measured
values of $x$ with the slopes being steeper than the those for
$F_2$ at the corresponding values of $x$, reflecting the dependence of
the cross section on the gluon distribution of the proton.
The features of the theoretical predictions discussed above
are again evident.  The description of the data is generally very good 
over the HERA kinematic range, with the CTEQ5F3 and MRST NNLO PDFs
providing the best description in the region of low $x$ and $Q^2$.

\subsubsection{Summary of Inclusive Charm Production}
The theoretical description of the
inclusive DIS charm cross section is seen to be generally 
very successful. A similar
picture has been observed for the differential $D^*$ meson
cross section data in both DIS~\cite{ZEUSDstar, H1Dstar, h1dstarjetsdis} and 
photoproduction~\cite{zeusdstarjetsgp,h1dstarjetsgp}
with the failures of the theoretical description confined to particular
corners of phase space.  The good description of the charm data is likely to
be a consequence of the fact that charm forms a
significant contribution of the total cross section
and is thus partly constrained by fits to the inclusive cross
section.
However, the increasing precision of the charm data
themselves mean that the range of theoretical predictions 
is now comparable to the spread of the experimental data.

\subsection{Beauty Production}
\label{beautyproduction}

\subsubsection{Experimental Techniques}
\label{bexptech}
The first measurement of beauty production in DIS was made
by ZEUS using the relative transverse momentum of muons relative to jets
($p_T^{\rm rel}$)~\cite{zeusBdis}.
The results were compared to NLO FFNS predictions using the parton 
level calculation from HVQDIS and hadronisation corrections 
estimated using Monte Carlo simulations.  The ZEUS data were found
to lie  around $2.5 \sigma$ above
the theoretical predictions. An analysis by H1~\cite{h1Bdis}, 
which combined the $p_T^{\rm rel}$ method with a
precise measurement of the displacement of the muon
track from the primary vertex using the silicon tracker,
produced a similar conclusion and the
data were found to lie $1.8 \sigma$ above the NLO predictions.  
The analysis technique based on the 
displacement of tracks due to the large lifetime of
beauty hadrons allowed a measurement of the inclusive
beauty DIS cross section at higher values of 
$Q^2$~\cite{H1f2cf2bhighq2,H1f2cf2blowq2}.  The
predictions of QCD were found to be in good agreement
with the data although the experimental errors
are somewhat larger when compared with the muon based analyses.

\subsubsection{Comparison of Experimental Beauty Data with PDFs}

\begin{figure}[htb]
 \begin{center}
\includegraphics[width=1.00\textwidth]{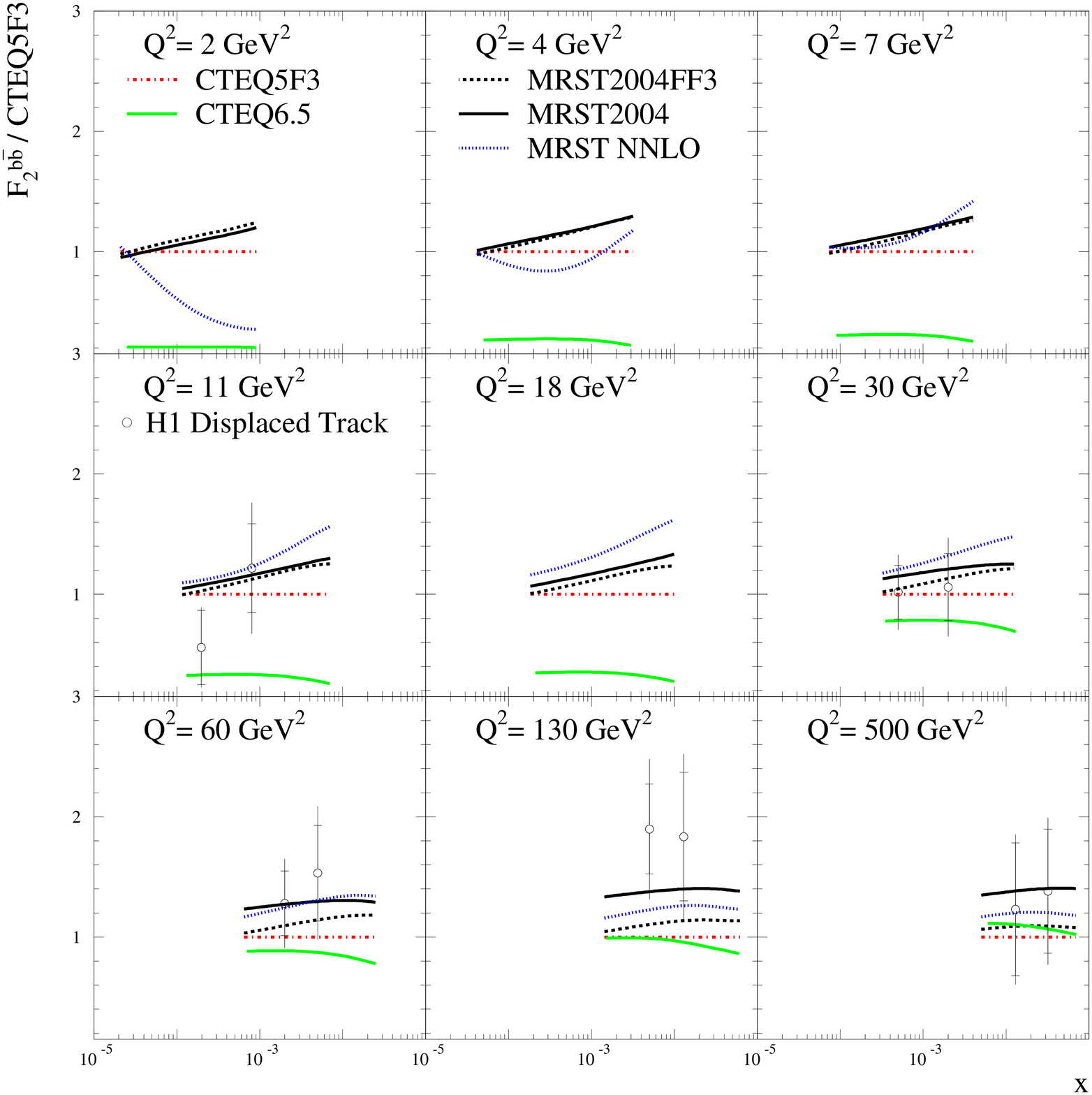}
   \caption{The ratio of the beauty structure function \ftwobb\ 
to the prediction of CTEQ5F3 as a function 
of $x$ for different values of $Q^2$.
The data obtained using the displacement of tracks from the primary vertex 
are shown as points. The inner error bar corresponds to the statistical
uncertainty and the outer error bar to the statistical and systematic
errors added in quadrature.  The data are compared with
the theoretical predictions of perturbative QCD using different prescriptions
for the heavy flavour treatment from the MRST and CTEQ fitting 
groups.}  
\label{fig:f2b}
\end{center}
\end{figure}

In figure~\ref{fig:f2b} the ratio of the beauty structure function \ftwobb\
to the FFNS prediction of CTEQ5F3 as a function of $x$ for different values 
of $Q^2$ using the same binning as for the studies of \ftwocc\ is shown.
The experimental data, obtained using the displaced track 
method~\cite{H1f2cf2bhighq2,H1f2cf2blowq2},
are shown as the points with error bars.   
The predictions for \ftwobb\ for the FFNS 
(CTEQ5F3, MRST2004FF3), the NLO VFNS (CTEQ6.5, MRST2004) and a
NNLO VFNS (MRST2004 NNLO) are also shown and discussed below.

For the two FFNS PDFs the
predictions for beauty production are very similar 
throughout the phase space, including the lowest values of $Q^2$, 
which is in contrast to the predictions for charm production.  
As will be shown in section~\ref{gluoncomp} this is a consequence of the 
beauty predictions accessing the gluon at higher values of 
the gluon momentum fraction $x_g$ 
where the differences
between the gluon distributions are smaller.

As was the case with the \ftwocc\ predictions the CTEQ VFNS PDF
(CTEQ6.5) differs considerably from the CTEQ FFNS PDF at low $Q^2$.
The difference is much larger for beauty than for charm with the 
VFNS prediction being $60-80\%$ lower than the FFNS prediction 
for $Q^2$ values up to $18 \ {\rm GeV^2}$.
The present experimental data at $Q^2 = 11 ~{\rm GeV}^2$
is compatible with both of the CTEQ predictions.
The MRST VFNS PDF is higher than that from CTEQ for all 
values of $Q^2$, with the difference still around $20\%$ at 
$Q^2 = 500 \ {\rm GeV^2}$.

As in the case of charm production 
the MRST FFNS and VFNS partons may be used to 
investigate
the difference in evolution of the two schemes.
The VFNS partons evolve quicker with increasing $Q^2$ than the FFNS partons
with the divergence continuing over the kinematic range shown, resulting
in a difference
of $30\%$ at $Q^2 = 500~{\rm GeV^2}$.

The NNLO VFNS PDF from MRST has a very different $x$ dependence and
normalisation when compared with the NLO VFNS at the lowest $Q^2$ values. 
The NNLO PDF then evolves much quicker than the NLO PDF being $20\%$
larger at $Q^2 = 18~{\rm GeV^2}$.  This is followed by a much slower evolution
such that the NNLO prediction is $20\%$ below the NLO prediction at
$Q^2 = 500~{\rm GeV^2}$.

It should be noted the predictions for inclusive beauty production using the 
ZM-VFNS are zero below $Q^2 = m_b^2 \sim 20 ~{\rm GeV}^2$. As can be seen in 
figure~\ref{fig:f2b} the region of phase space below the `mass threshold' is 
experimentally accessible. The measurements of beauty production significant
from zero at $Q^2 = 11~{\rm GeV^2}$ highlight the fact that the ZM-VFNS is
inapplicable for the description of 
heavy flavour production in the kinematic range of HERA.

\subsubsection{Summary of Inclusive Beauty Production}

The inclusive beauty data are well described by all of the
theoretical predictions, within the large statistical errors, with
no indication of the excess over theory as observed in the final state
measurements. The differences between any set of theoretical predictions 
are generally smaller for beauty production than they are for charm due to 
the region of $x_gg(x)$ probed. 
However, in the region $Q^2 < m_b^2$ there are much larger differences 
between the theoretical predictions.
One possibility for the large spread in the predictions 
in this region is that beauty is a small contribution to the total cross 
section and the experimental uncertainties on the beauty cross
section measurements are still large.
The large beauty mass means that, in contrast to charm, $Q^2$ values well 
below $Q^2 = M^2$ but still in a region where pQCD may be expected to work
are experimentally accessible.

\section{Theoretical Uncertainties}
\label{theoryuncert}
The theoretical uncertainty due to the effects of missing higher order
corrections on QCD predictions performed at a fixed order is usually 
estimated by varying the factorization and renormalization scales
in the calculation.  
The precise prescription
adopted varies considerably.  This is particularly true for heavy flavours
where there are further degrees of freedom, in comparison to the total 
inclusive cross section, due to the mass of the heavy quarks.  It should be noted 
that in kinematic 
regions where the chosen QCD scheme is a poor approximation 
the variation of the 
perturbative scales does not provide an estimate of the effect 
of the neglected contributions. An illustrative example
of this is the ZM-VFNS for values of  $Q^2$ below the heavy flavour
`threshold' $Q^2 = M^2$ where the cross section prediction is zero and the
variation of the perturbative scale has no effect.

\begin{figure}[htb]
 \begin{center}
\includegraphics[width=1.00\textwidth]{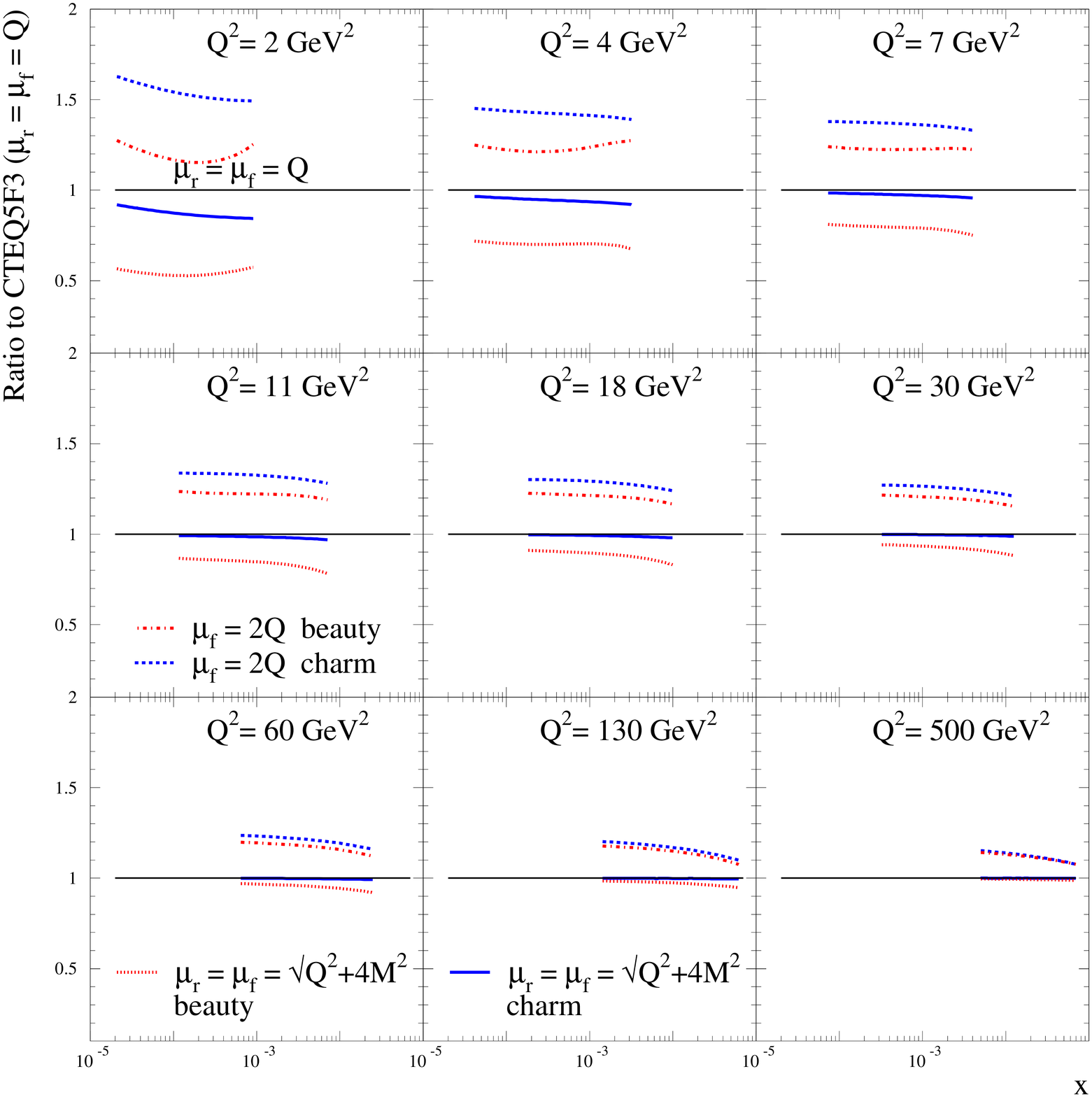}
   \caption{The effect of varying the renormalization 
and factorization scales for the inclusive
charm and beauty DIS cross section shown as
a function of $x$ for different values of $Q^2$.
The predictions are shown using the 
CTEQ5F3 PDF with the default 
scale $\mu_r = \mu_f = Q$ and alternatively with 
$\mu_r = \mu_f/2 = Q$ and  $\mu_r = \mu_f = \sqrt{Q^2+M^2}$.
}  
\label{fig:f2ratio}
\end{center}
\end{figure}

In this section a comparison of the size of the theoretical uncertainty on 
inclusive charm and beauty cross sections is investigated
using the FFNS (CTEQ5F3) as an example. In figure~\ref{fig:f2ratio}
the uncertainty on \ftwocc\ and \ftwobb\ due to the 
variation of the factorization
and renormalization scales is shown.  The default scales are chosen
as $\mu_f = \mu_r = Q$.

In the figure the variation of the factorization
scale $\mu_f$ from $Q$ to $2Q$ leads to large increases in the predictions,
particularly at low $Q^2$. The uncertainty due to varying $\mu_f$ for charm 
is significantly larger than for beauty
at the lowest values of $Q^2$, and arises from the difference in the
region of $x_g$ probed (see section~\ref{gluoncomp}).  The variation
of the factorization scale from $Q$ to $Q/2$ (not shown) leads to 
decreases in the predictions of similar magnitude.

The choice of $Q$ as the renormalization scale means that
the variation of $\mu_r$ from $Q$ to $2Q$($Q/2$) leads to 
similar effects for charm and beauty (not shown).
The uncertainty due to these changes is extremely large at the lowest value 
of $Q^2$ $-50\%$($+500\%$)
and decreases with increasing $Q^2$ to around $-20\%$($+30\%$) 
for the highest 
$Q^2$ value shown. The variations
reflect the behaviour of the strong coupling $\alpha_s(\mu_r)$
with the scale $\mu_r$.

For predictions of final state cross sections
larger perturbative scales than $Q^2$ are often used, 
typically $\mu_f = \mu_r = \sqrt{Q^2+4M^2}$~\cite{hera-lhc},
which provide a better estimation of the virtuality of the hard scattering
process.
The relative changes to the predictions for the scale choice 
$\mu_f = \mu_r = \sqrt{Q^2+4M^2}$ are also shown in the figure.
Due to the large mass of the beauty quark
there is a large suppression of the cross section, particularly at low $Q^2$.
The effect is much smaller for charm production.  For the
scale choice $\mu_f = \mu_r = \sqrt{Q^2+4M^2}$ the variation of
$\mu_f$ and $\mu_r$ by factors $2$ and $1/2$ leads to improved stability in the
predictions than those observed for the 
scale choice $\mu_f = \mu_r =Q$. For example, the variation of $\mu_r$
by $0.5$ leads to maximum changes of $+70\%$($+30\%$) for charm (beauty)
and the variation of  $\mu_f$ by factors $0.5$ and $2$ leads to 
maximum changes of $\pm 40\%$ for charm and $\pm 20\%$ for beauty.

At higher values of $Q^2$, where the role of the heavy quark
masses in the perturbative scale becomes decreasingly important, the 
uncertainty on the
charm and beauty cross sections becomes similar.  At values of 
$Q^2 = 500~{\rm GeV^2}$
the combined theoretical uncertainty is around $20\%$. This is to be compared
with the differences between the predictions of the MRST FFNS and VFNS partons 
of $\sim 20\%$ for charm and $\sim 30\%$ for beauty at $Q^2 = 500~{\rm GeV^2}$.

In summary the uncertainty due to variations of the scale are
large, particularly at low $Q^2$. The large uncertainty from variation of the
scales when using $\mu_f = \mu_r =Q$ is reduced by using the larger 
scale $\mu_f = \mu_r = \sqrt{Q^2+4M^2}$.
The fact that charm production gives a significant contribution to the 
total cross section suggests that the scale uncertainty may be
overestimated in this manner i.e. the inclusive DIS cross section will
be poorly described with the changed charm scale.  To compensate for this
effect, the PDFs given in~\cite{ccfmfits} 
based on CCFM evolution, are available for
a number of different choices of scale for the heavy flavour contribution, 
which reduces the scale uncertainty when evaluated as in the manner above.  

\section{Comparison of the CTEQ and MRST Gluon Distributions}
\label{gluoncomp}
\begin{figure}[htb]
 \begin{center}
\includegraphics[width=1.00\textwidth]{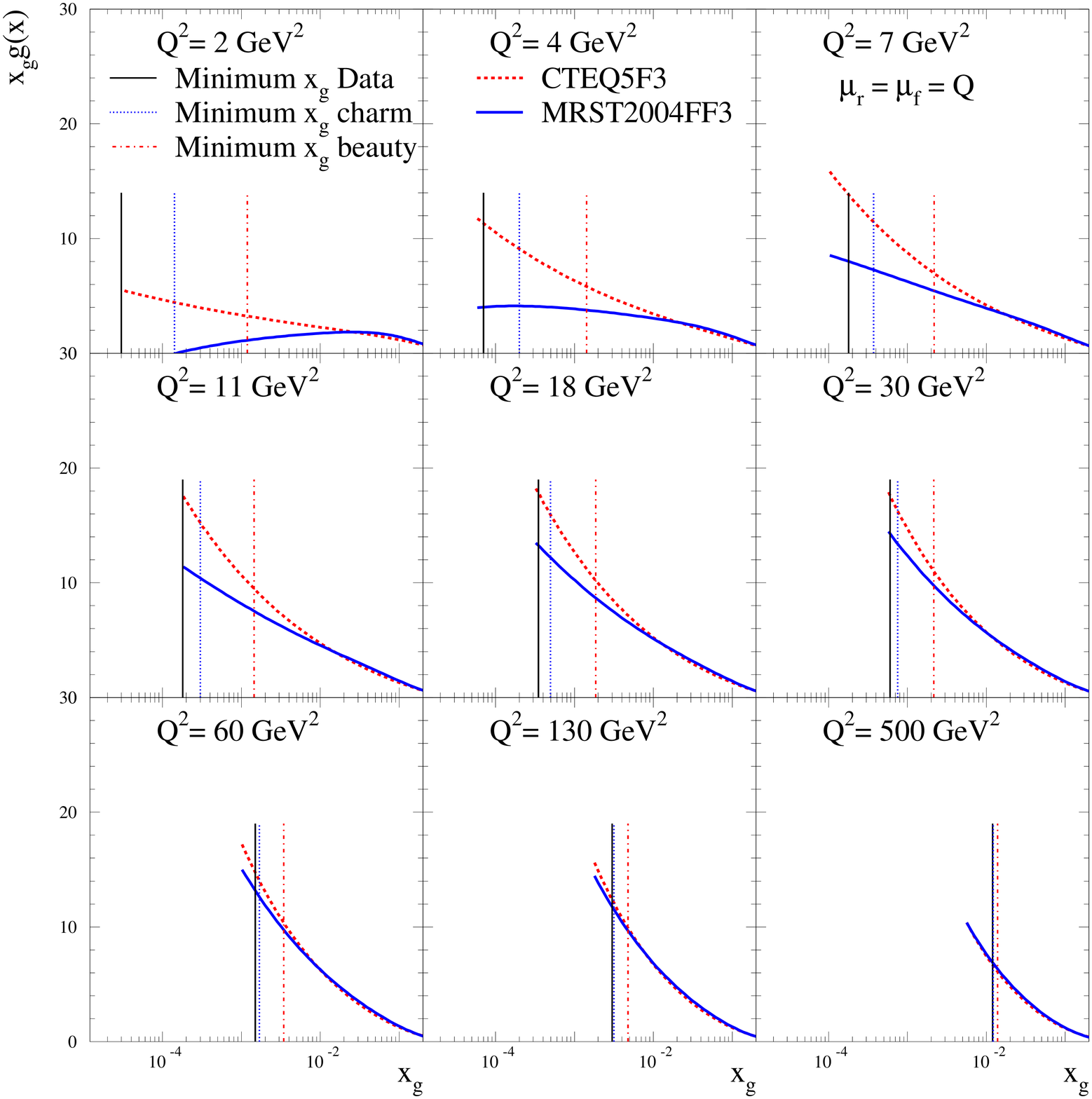}
   \caption{
The CTEQ and MRST gluon distributions shown as
a function of $x_g$ for 
different values of $\mu^2 = Q^2$. The vertical solid lines indicate
the range of experimental \ftwocc\ data. The vertical
dashed and, dashed-dotted lines indicate the effective range
probed for charm and beauty production, respectively.
}  
\label{fig:gluonrange}
\end{center}
\end{figure}

In this section the differences in the CTEQ and MRST gluon distributions 
at low $Q^2$ and $x$ are highlighted since they have an important effect on the
predictions for the inclusive charm cross section. At leading order
the cross section for heavy flavour production in the FFNS is given by the 
convolution $\sigma \propto \int_{ax}^1 C(x/x_g,\mu^2)g(x_g) dx_g/x_g$
where $C(x/x_g,\mu^2)$ is the coefficient function and 
$g(x_g)$ is the gluon distribution. The integral is calculated over the 
range from threshold values
of $x_g=ax$ where $a= 1+4M^2/Q^2$ up to a maximum value of $x_g=1$. 
The coefficient functions have the feature that the average value of $x_g$ is 
weighted away from $x=x_g$ and the convolution results in probing the gluon
distribution at higher values of 
$x_g$\footnote{At NLO the weighting away from $x=x_g$
in the coefficient function increases.}.  
Therefore, at low values of $Q^2$ 
the minimum value of $x_g$ probed ($x_g^{\rm min}= ax$) is higher for beauty 
production than charm. 
This is 
illustrated in figure~\ref{fig:gluonrange} where the CTEQ and MRST 
FFNS gluon distributions
are compared as a function of $x_g$ at different values of  $\mu^2 = Q^2$.  
For each bin the minimum $x_g$ value of the 
experimental \ftwocc\ charm data is 
indicated as well as the 
minimum $x_g$ value for charm and beauty production.  
As can be seen in the figure,
the charm data is more sensitive to the 
large difference in the CTEQ and MRST FFNS gluon distributions at low $Q^2$ 
and $x_g$.   
The differences in the gluon distributions contribute
to the differences in the charm cross section between CTEQ5F3 and MRST2004FF3
seen in the lowest $Q^2$ regions of figure~\ref{fig:f2c}. 
The difference in the charm cross section for the
two PDFs is smaller than the difference in the gluon PDFs because of the 
convolution
of the gluon distribution with the heavy flavour coefficients.  
The similarity of the CTEQ and MRST FFNS
gluon distributions for the $x_g$ range sensitive to beauty production 
explains 
why the large differences at low $Q^2$ between CTEQ5F3 and MRST2004FF3
observed for \ftwocc\ are not evident for \ftwobb\ (figure~\ref{fig:f2b}).

\section{Summary and Outlook}

In this paper the latest
theoretical predictions for heavy flavour production in DIS
have been introduced
and the status of the comparison with the experimental data from HERA
investigated.  
The inclusive charm data, which form a sizeable fraction of
the total cross section, are seen to be generally well
described by the predictions. 
However, the increasing precision of the charm data
mean that the range of theoretical predictions 
is now comparable to the spread of the experimental data.
The production of charm
is particularly sensitive to the gluon distribution of the
proton. Future measurements of the inclusive charm cross
section from the HERA-II data 
will continue to provide tests of the theoretical models and 
PDFs.

The inclusive production of beauty is also well described by the 
QCD predictions. The production of beauty in the region 
$Q^2 < m_b^2 \sim 20~{\rm GeV^2}$
is particularly sensitive to the scheme and perturbative
scale chosen.  The theoretical understanding will clearly benefit
from the increased statistics in this region
that will become available from the HERA II data. 

The theoretical predictions for charm and beauty show sensitivity in the
$Q^2$ dependence due to different evolution schemes, although the present
theoretical picture is complicated by mass threshold and other effects.
The sensitivity of heavy flavour production to the parton distributions 
of the proton will continue to improve the reliability of QCD predictions
of standard model and new physics processes at the Tevatron and LHC.

\section*{Acknowledgements}

I am grateful to R.~S.~Thorne and W.~K.~Tung for providing me with their 
calculations and for productive discussions.


\newpage

\begin{thebibliography}{99}



\bibitem{H1f2cf2bhighq2}
  A.~Aktas {\it et al.}  [H1 Collaboration],
  Eur.\ Phys.\ J.\ C {\bf 40}  (2005) 349 
  [hep-ex/0411046].


\bibitem{H1f2cf2blowq2}
 A.~Aktas {\it et al.}  [H1 Collaboration],
  Eur.\ Phys.\ J.\ C {\bf 45} (2006) 23
  [hep-ex/0507081].



\bibitem{H1Dstar94}
C.~Adloff {\it et al.}  [H1 Collaboration],
Z.\ Phys.\ C {\bf 72} (1996) 593
  [hep-ex/9607012].


\bibitem{ZEUSDstar97}
J.~Breitweg {\it et al.}  [ZEUS Collaboration],
Eur.\ Phys.\ J.\ C {\bf 12} (2000) 35
  [hep-ex/9908012].



\bibitem{H1Dstar}
C.~Adloff {\it et al.}  [H1 Collaboration],
Phys.\ Lett.\ B {\bf 528} (2002) 199
  [hep-ex/0108039].


\bibitem{ZEUSDstar}
S.~Chekanov {\it et al.}  [ZEUS Collaboration],
Phys.\ Rev.\ D {\bf 69} (2004) 012004
  [hep-ex/0308068].




\bibitem{Cteq5}
H.~L.~Lai {\it et al.}  [CTEQ Collaboration],
Eur.\ Phys.\ J.\ C {\bf 12} (2000) 375
  [hep-ph/9903282].


\bibitem{Cteq65}
W.~K.~Tung, H.~L.~Lai, A.~Belyaev, J.~Pumplin, D.~Stump and C.~P.~Yuan,
  [hep-ph/0611254].



\bibitem{Martin:2004dh}
A.~D.~Martin, R.~G.~Roberts, W.~J.~Stirling and R.~S.~Thorne,
Eur.\ Phys.\ J.\ C {\bf 39} (2005) 155
  [hep-ph/0411040].

\bibitem{ThorneNNLO}
  R.~S.~Thorne,
  Phys.\ Rev.\ D {\bf 73} (2006) 054019
  [hep-ph/0601245].


\bibitem{mrstff}
A.~D.~Martin, W.~J.~Stirling and R.~S.~Thorne,
Phys.\ Lett.\ B {\bf 636} (2006) 259
 [hep-ph/0603143].




%
%

\bibitem{ffnsnlo}
E.~Laenen, S.~Riemersma, J.~Smith and W.~L.~van Neerven,
Nucl.\ Phys.\ B {\bf 392} (1993) 162;
%
E.~Laenen, S.~Riemersma, J.~Smith and W.~L.~van Neerven,
Nucl.\ Phys.\ B {\bf 392} (1993) 229.



\bibitem{HVQDIS}
B.~W.~Harris and J.~Smith,
Nucl.\ Phys.\ B {\bf 452} (1995) 109
  [hep-ph/9503484].



\bibitem{h1lowq2}
  C.~Adloff {\it et al.}  [H1 Collaboration],
  Eur.\ Phys.\ J.\ C {\bf 21} (2001) 33
  [hep-ex/0012053].

\bibitem{zeusqcd}
  S.~Chekanov {\it et al.}  [ZEUS Collaboration],
  Phys.\ Rev.\ D {\bf 67} (2003) 012007
  [hep-ex/0208023].


\bibitem{diffcharm}
  A.~Aktas {\it et al.}  [H1 Collaboration],
  [hep-ex/0610076].


\bibitem{VFNS2}
R.~S.~Thorne and R.~G.~Roberts,
Phys.\ Rev.\ D {\bf 57} (1998) 6871
  [hep-ph/9709442];


R.~S.~Thorne and R.~G.~Roberts,
Phys.\ Lett.\ B {\bf 421} (1998) 303
  [hep-ph/9711223];

R.~S.~Thorne and R.~G.~Roberts,
Eur.\ Phys.\ J.\ C {\bf 19} (2001) 339
  [hep-ph/0010344].


\bibitem{zeusjets}
  S.~Chekanov {\it et al.}  [ZEUS Collaboration],
  Eur.\ Phys.\ J.\ C {\bf 42} (2005) 1
  [hep-ph/0503274].




\bibitem{ACOT}
  J.~C.~Collins and W.~K.~Tung,
Nucl.\ Phys.\ B {\bf 278} (1986) 934.

M.~A.~G.~Aivazis, F.~I.~Olness and W.~K.~Tung,
Phys.\ Rev.\ D {\bf 50} (1994) 3085
  [hep-ph/9312318];

M.~A.~G.~Aivazis, J.~C.~Collins, F.~I.~Olness and W.~K.~Tung,
Phys.\ Rev.\ D {\bf 50} (1994) 3102
  [hep-ph/9312319];

J.~C.~Collins,
Phys.\ Rev.\ D {\bf 58} (1998) 094002
  [hep-ph/9806259].


\bibitem{ACOTchi}
W.~K.~Tung, S.~Kretzer and C.~Schmidt,
J.\ Phys.\ G {\bf 28} (2002) 983
  [hep-ph/0110247].



\bibitem{SACOT}
M.~Kr\"{a}mer, F.~I.~Olness and D.~E.~Soper,
Phys.\ Rev.\ D {\bf 62} (2000) 096007
  [hep-ph/0003035].




\bibitem{PDG2006}
W.~M.~Yao {\it et al.}  [Particle Data Group],
J.\ Phys.\ G {\bf 33} (2006) 1.


\bibitem{hera-lhc}
  S.~Alekhin {\it et al.},
``HERA and the LHC - A workshop on the implications of HERA for LHC  physics:
Proceedings Part B,'' p 405.
  [hep-ph/0601013].



\bibitem{peterson}
C.~Peterson, D.~Schlatter, I.~Schmitt and P.M.~Zerwas,
Phys.\,Rev.\ D {\bf 27} (1983) 105.



\bibitem{Chuvakin:2000zj}
  A.~Chuvakin, J.~Smith and B.~W.~Harris,
  Eur.\ Phys.\ J.\ C {\bf 18} (2001) 547
  [hep-ph/0010350].




\bibitem{ccfm2}
M.~Ciafaloni,
Nucl.\ Phys.\ B {\bf 296} (1988) 49;

S.~Catani, F.~Fiorani and G.~Marchesini,
Phys.\ Lett.\ B {\bf 234} (1990) 339;

S.~Catani, F.~Fiorani and G.~Marchesini,
Nucl.\ Phys.\ B {\bf 336} (1990) 18;

G.~Marchesini,
Nucl.\ Phys.\ B {\bf 445} (1995) 49
 [hep-ph/9412327].



\bibitem{h1dstarjetsdis}
  A.~Aktas {\it et al.} [H1 Collaboration],
  [hep-ex/0701023].




\bibitem{zeusdstarjetsgp}
  J.~Breitweg {\it et al.}  [ZEUS Collaboration],
  Eur.\ Phys.\ J.\  C {\bf 6} (1999) 67
  [hep-ex/9807008];

  S.~Chekanov {\it et al.}  [ZEUS Collaboration],
  Phys.\ Lett.\  B {\bf 565} (2003) 87
  [hep-ex/0302025];

  S.~Chekanov {\it et al.}  [ZEUS Collaboration],
  Nucl.\ Phys.\ B {\bf 729} (2005) 492
  [hep-ex/0507089].


\bibitem{h1dstarjetsgp}
  A.~Aktas {\it et al.}  [H1 Collaboration],
  [hep-ex/0608042].


\bibitem{zeusBdis}
S.~Chekanov {\it et al.}  [ZEUS Collaboration],
Phys.\ Lett.\ B {\bf 599} (2004) 173
  [hep-ex/0405069].


\bibitem{h1Bdis}
  A.~Aktas {\it et al.}  [H1 Collaboration],
  Eur.\ Phys.\ J.\ C {\bf 41} (2005) 453
  [hep-ex/0502010].





\bibitem{ccfmfits}
  H.~Jung,
  [hep-ph/0411287].





\end{thebibliography}
\end{document}